\documentclass[screen]{acmart}

\AtBeginDocument{%
  }

\copyrightyear{2025}
\acmYear{2025}
\setcopyright{rightsretained}
\acmConference[LAK 2025]{LAK25: The 15th International Learning Analytics and Knowledge Conference}{March 03--07, 2025}{Dublin, Ireland}
\acmBooktitle{LAK25: The 15th International Learning Analytics and Knowledge Conference (LAK 2025), March 03--07, 2025, Dublin, Ireland}
\acmPrice{}
\acmDOI{10.1145/3706468.3706487}
\acmISBN{979-8-4007-0701-8/25/03}

\begin{document}


\title{A Novel Approach to Scalable and Automatic Topic-Controlled Question Generation in Education}


\author{Ziqing Li}
\affiliation{%
  \institution{Department of Computer Science, University College London}
  \city{London}
  \country{United Kingdom}}
\email{ziqing.li.19@ucl.ac.uk}

\author{Mutlu Cukurova}
\affiliation{%
  \institution{UCL Knowledge Lab, University College London}
  \city{London}
  \country{United Kingdom}}
\email{m.cukurova@ucl.ac.uk}

\author{Sahan Bulathwela}
\affiliation{%
  \institution{Centre for Artificial Intelligence, University College London}
  \city{London}
  \country{United Kingdom}}
\email{m.bulathwela@ucl.ac.uk}

\renewcommand{\shortauthors}{Li et al.}

\begin{abstract}
The development of Automatic Question Generation (QG) models has the potential to significantly improve educational practices by reducing the teacher workload associated with creating educational content. This paper introduces a novel approach to educational question generation that controls the topical focus of questions. The proposed Topic-Controlled Question Generation (T-CQG) method enhances the relevance and effectiveness of the generated content for educational purposes. Our approach uses fine-tuning on a pre-trained T5-small model, employing specially created datasets tailored to educational needs. The research further explores the impacts of pre-training strategies, quantisation, and data augmentation on the model's performance. We specifically address the challenge of generating semantically aligned questions with paragraph-level contexts, thereby improving the topic specificity of the generated questions. In addition, we introduce and explore novel evaluation methods to assess the topical relatedness of the generated questions. Our results, validated through rigorous offline and human-backed evaluations, demonstrate that the proposed models effectively generate high-quality, topic-focused questions. These models have the potential to reduce teacher workload and support personalised tutoring systems by serving as bespoke question generators. With its relatively small number of parameters, the proposals not only advance the capabilities of question generation models for handling specific educational topics but also offer a scalable solution that reduces infrastructure costs. This scalability makes them feasible for widespread use in education without reliance on proprietary large language models like ChatGPT.

\end{abstract}



\received{Not Available}
\received[revised]{Not Available}
\received[accepted]{Not Available}

\maketitle

\section{Introduction}
One of the most significant and pertinent challenges facing education systems today is the teachers' workload. It is argued to be the main reason behind issues associated with teachers' retention in the profession as well as the lack of interest among graduate students to go into teaching professions. On the other hand, according to the Global Report on Teachers published by the Teacher Task Force and UNESCO, 44 million additional teachers will be needed by 2030 to meet Sustainable Development Goal 4 (SDG4), which aims to achieve universal primary and secondary education for all \cite{unesco2024global}, without any improvements to the \textit{status quo}. Among these concerns, generative AI in education is seen as an opportunity to 'transform a teacher’s day-to-day work' \cite{gov2024use} by reducing their workload and improving educational outcomes through the automation of routine tasks.  

Creating lesson materials and generating topic-specific, relevant, and age-appropriate questions for teaching have long been identified as time-intensive tasks for teachers, and an area where increased consistency is also expected to improve educational outcomes for students \cite{giannakos2024promise}. Although learning analytics and AI in Education researchers have long explored ways to support teachers' question generation capabilities through data-driven insights and models, attempts on \textbf{Topic-Controlled Question Generation (T-CQG)} have been less successful, primarily due to the lack of quality in the generated content. The use of large language models (LLMs) in teacher-facing interfaces, however, has the potential to address these quality concerns by leveraging recent advancements in NLP for automatic educational question generation (EdQG). EdQG can help teachers reduce the labor-intensive task of generating questions to promote classroom discussions, design formative and summative assessments, create lesson hooks, or address student misconceptions which are all activities that teachers consider among the most time-consuming in their profession \cite{gov2024use}. Although most issues related to teachers' workload are complex, ecosystem-level socio-technical challenges \cite{cukurova2023adoption}, T-CQG can serve as a small yet important practical step towards enhancing teachers' productivity, aiming to mitigate workload and address issues related to teacher retention and attraction to the profession.

In addition to their potential to support teachers, EdQG (and T-CQG) models can be integrated into learning management systems (LMSs) and intelligent tutoring systems (ITSs), to advance the system's capability to perform precise diagnostics on learner's knowledge gaps. The responses received from learners can inform the learning analytics pipeline more precisely and frequently to have a refined learner state representation, that can empower the system with targeted interventions. However, such interventions require advancements to generic neural network question generation models that do not have the ability to contextualise generation with constraints.


{The novelties that we introduce through this work are three-fold. We 1) propose a novel method to generate a dataset with contrastive examples in order to effectively train a T-CQG model and 2) validate and propose novel ways of evaluating the topical relatedness of the generations to the controlled topic using semantic relatedness metrics while 3) this is the only work that attempts in using a very small language model (sLM) with $\approx 60M$ parameters,  and succeeds in producing a T-CQG neural model.}

\section{Problem Definition, Background Research, and Research Questions} \label{sec:problem}

In this section, we introduce the formal problem definition and prior work, leading to the research questions.

\subsection{Problem Definition}\label{sec:problemdef}
Although language models have been employed for question generation, their application in educational settings has only recently begun to be systematically explored with a heavy focus on the potential practical applications of proprietary models (e.g. GPT models' prompt engineering and RAG applications for question generation). While existing research in relevant academic communities with a more technical focus explores generating questions from descriptive texts \cite{Du2017,wang2018qgnet}, the task remains highly complex and there is less focus on the educational value of the generated questions in evaluations. Context plays a crucial role in the educational value of EdQG, yet much existing work has focused primarily on generating questions from sentences, paragraphs, or structured data in isolation \cite{hu2018topic, lopez2021simplifying}, with limited attention given to topic-controlled question generation in a given context. 

Topic-controlled question generation takes a target topic in addition to the descriptive text as context into account while generating the models' outputs. On the other hand, traditional approaches which take sentences or paragraphs as inputs without contextual topic-control tend to generate questions that arbitrarily combine or select concepts and topics which are likely to be of limited practical value to professionals like teachers. From the learners' points of view, prior research also suggests a strong correlation between the personalisation of testing and knowledge retention \cite{Bahrick1993}, which further supports the importance of topic-controlled question generation. Developing comprehensive, high quality and relevant educational question sets across different topics can significantly enhance teaching practice and support students through intelligent tutoring systems that provide personalised learning to diverse learners. 

In the scope of this work, we define topic-controlled question generation (T-CQG) as follows. Let us suppose a learner $\ell$ has already consumed learning materials that contain the knowledge context $c$ containing various topics $T_c$. A goal of a teacher or an intelligent system is then to generate a question $\hat{q}_t$, where $\hat{q}_t$ is an educational question about the target topic $t$, where $t \in T_c$, and $\hat{q}_t$ consists of a sequence of tokens $q_t \in \{w_1, \dots, w_{|q_t|}\}$ of arbitrary length $|q_t|$. This task requires that the question is not only contextually relevant to the paragraph context ${c}$, but also closely aligned with the thematic focus defined by topic $t$. The probability $p(q_t | c,t)$ incorporates the coherence and relevance of each token in the sequence, rendering the generation process highly sensitive to both the context and the topic. This task can be mathematically defined to identify the optimal question $\hat{q_t}$ that maximises the conditional probability as per equation \ref{eq:problem}.

\begin{equation} \label{eq:problem}
\hat{q_t} = \arg \max_{q_t} p({q_t} | {c},{t}) = \arg \max_{q_t} \sum_{i=1}^{|{q_t}|} \log p(w_{i} | {c}, {t}, w_1 \dots w_{i-1})    
\end{equation}

where, $p({q_t} | {c},{t})$ denotes the conditional probability that also depends on the tokens $w \in q_t$.


\subsection{Related Work}


Question Generation (QG) involves automatically generating questions from a specific text passage or a document. The main goal of QG is to produce questions that are not only syntactically and semantically correct but also contextually relevant and meaningful for the intended use. There has been a growing use of computational models to generate contextually relevant and grammatically correct questions \cite{wang2024llmseducation}. In educational contexts specifically, QG has been implemented in various systems including intelligent tutoring systems \cite{yadav2023cp}, writing support systems \cite{pinto2023write}, and knowledge assessment platforms \cite{kuo2023knowledge}.

Existing research categorises QG into two types: answer-aware and answer-agnostic \cite{zhang2021review}. In \emph{answer-aware} QG, the target answer is predetermined, and questions are generated to correspond with this answer within the given text context. On the other hand, \emph{answer-agnostic} QG does not provide the target answer to the language model, allowing for more open-ended question generation which are considered to be educationally more valuable. However, answer-agnostic QG is a more challenging task for NLP research. Early research in answer-agnostic QG relied heavily on rule-based techniques that required experienced educators to develop rules that could convert declarative sentences into interrogative forms \cite{HeilmanSmith2010, Adamson2013}. These methods, while effective, are labour-intensive and time-consuming, demanding significant manual effort in creating high-quality, handcrafted rules \cite{Chen2021}, which inherently limits their scalability and diversity in question generation. These limitations led more recent research investigations to focus on data-driven neural network (NN) approaches.  

Early implementations of QG practices with data-driven approaches predominantly utilized sequence-to-sequence (seq2seq) architectures incorporating Recurrent Neural Networks (RNNs) \cite{Du2017}. More recently, the focus shifted towards employing end-to-end techniques facilitated by deep neural networks \cite{zhang2021review}. For instance, \cite{dathathri2020plug} and \cite{khalifa2021distributional} utilised GPT-2 combined with either an attribute classifier or training another autoregressive language model to guide the generated text towards a topic. However, these approaches typically generated content that is too broadly categorised (such as a category being 'science'), failing to achieve the level of topic specificity required for them to be of real-world value for educational practitioners. On the other hand, a more targeted approach by \cite{hu2018topic} employing an LSTM model equipped with a pre-decoding mechanism, demonstrated the ability to generate questions on detailed topics. This model, though promising in its specificity, was only applied at the sentence level, limiting its utility for broader educational applications. More contemporary models leverage pre-trained transformers like GPT \cite{blobsteinangel,elkins2024teachers} (Decoder Only) and T5 (Text-to-Text Transfer Transformer) \cite{vachev2022leaf,bulathwela2023scalable} (Endoder-Decoder). These advanced NLP approaches like transformer architectures have shown to be effective in generating coherent and relevant questions {for specified texts \cite{AlFaraby2024}.} However, their use in meaningful and relevant educational question generation needs further explorations and evaluations in educational contexts \cite{bulathwela2023scalable,vachev2022leaf}. In short, the problem of topic-specific question generation as scoped in section \ref{sec:problemdef} has been of interest to multiple researchers in the past, yet it is still an open challenge for the community.

One of the significant challenges in research utilising pre-trained transformer architectures for the scoped problem is the issue of making generated content more specifically aligned with the particular topics studied and its contextual considerations. Previous literature in AI in Education research proposed multiple approaches when linking knowledge components of topics to generated learning materials such as questions. The most common approach is expert human labelling, but it is challenging to be scaled even though its accuracy is unmatched \cite{Yudelson13} and considered as gold-standard. Due to the scaling challenges of expert human labelling, recent works have also proposed methods such as entity linking \cite{Brank2017,Tagme} that provide scalability even if it tends to sacrifice some accuracy. Another proposed approach is the so-called "Wikification" which is the practice of using Wikipedia as a source for semantic annotations \cite{ZhangRettinger2014}. The approach has demonstrated significant advancements in recent years and offers considerable potential for automatically extracting concepts from Wikipedia entries to generate topic-specific educational materials \cite{bulathwela2021semantic}. Additionally, since there has been extensive research on Wikipedia for its potential for semantic labelling of AI-generated content, its concept relatedness metrics that are based on its link structure, page co-occurrence etc. \cite{Ponza2020} are well developed and can represent semantic relatedness between Wikipedia concepts to a high accuracy. However, the use of approaches that allow scalable solutions such as Wikification \cite{Brank2017} in educational question generation models is yet to be explored in detail in learning analytics.

\subsection{Research Questions}

This paper aims to address these challenges associated with the topic-controlled EdGQ. We conducted supervised fine-tuning on a pre-trained T5-small model (hereafter referred to as the T5 model), an approach that is preferable and safer for educational entities to manage and control the language model (LM) with minimal infrastructure costs. The fine-tuning process utilised the novel MixSQuAD dataset, an enrichment of the SQuAD dataset \cite{rajpurkar2016squad}, which is a commonly used general question generation dataset characterised by its shallow questions. Additionally, we designed experiments to explore the impacts of pre-training strategies, text data augmentation, and model quantisation on the model’s performance. We evaluated the model on the novel MixKhanQ dataset, derived based on the KhanQ dataset  \cite{gong-etal-2022-khanq}, which features human-like, in-depth questions sourced from Khan Academy, an online education platform. This is designed to assess the model’s effectiveness on academic materials, and its ability to generate educationally meaningful questions to explore its practical value for teaching and learning contexts. Based on these steps, the paper proposes a novel set of models that can perform high-precision topic-controlled educational question generation (T-CQG). The research questions addressed through this work are as follows: 

\begin{itemize}
    \item \textbf{RQ1:} What are the most representative metrics for automated measures of generated questions on their topic relevance considering human evaluations as the ground truth?
    \item \textbf{RQ2:} Is it feasible to fine-tune a pre-trained language model (PLM) to perform T-CQG?
    \item \textbf{RQ3:} Can further pre-training of the PLM on scientific text improve the quality of T-CQG? 
    \item \textbf{RQ4:} How does model quantisation affect the performance of the fine-tuned models while improving scalability?
    \item \textbf{RQ5:} To what extent can data augmentation further improve the quality of T-CQG?

    
    
\end{itemize}

 





\section{Methodology}\label{methodology} 

\subsection{Datasets Utilised}\label{datasets_utilised}


We used the SQuAD 1.1, the Stanford Question Answering Dataset, comprising over 100,000 questions crafted by crowd workers based on a selection of 536 Wikipedia articles \cite{rajpurkar2016squad} as the source for creating new datasets (SQuAD+, MixSQuAD and MixSQuAD2X as described in section \ref{sec:noveldata} below) for finetuning the models. When training the TopicQGedu Model (see section \ref{topicQedu} below), we used PeS2O dataset \cite{peS2o}, a collection of scientific abstracts, to perform the pre-training as prior work has shown this may increase the model's performance in educational settings \cite{bulathwela2023scalable}.

For evaluation, we used the KhanQ dataset \cite{gong-etal-2022-khanq} as it presents a more relevant challenge for educational question generation. It includes 1,034 high-quality questions in the STEM fields generated by learners, which aim to probe deep understanding of subjects taught in Khan Academy's online courses \footnote{\url{https://www.khanacademy.org}}. Despite its smaller size relative to SQuAD, KhanQ aligns more closely with our objective to generate topic-based and relevant educational questions (as per prior work \cite{fawzi2024humanlike}). To adapt the dataset for topic-based evaluation, we use the same approach as MixSQuAD (section \ref{sec:mixsquad}) to create a dataset with contrasting topic-based questions. We refer to the transformed version of the KhanQ dataset as \emph{MixKhanQ} dataset. 



\subsection{Creating Novel Datasets for T-CQG}\label{sec:noveldata} 

\begin{figure*}[h]
\begin{center}
\includegraphics[width=.63\linewidth]{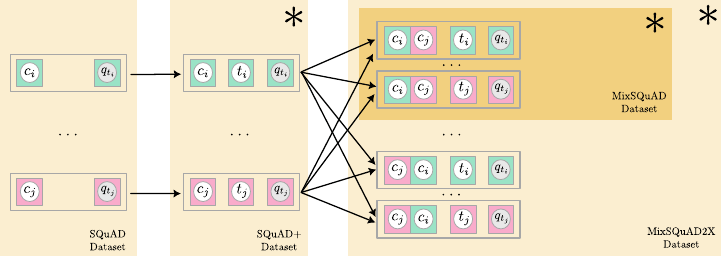}
\caption{Methodology for generating the different training datasets proposed in the model from the contexts $c$, topics $t$ and target questions $q_t$ (shaded as label) from the SQuAD dataset is illustrated using two random examples from the dataset, example $i$ (green) and example $j$ (pink). The contexts $c_i$ and $c_j$ glued together are concatenated texts treated as a single field in the dataset. The orange rectangles indicate the scope of the datasets while (*) marks the newly proposed datasets.}
\label{fig:datasets}
\end{center}
\end{figure*}

A core contribution of this work is to introduce a novel data enrichment method that leads to the creation of new datasets that are derived from conventional question generation datasets. As described in \ref{datasets_utilised}, we derive the new datasets from SQuAD and KhanQ. These datasets already contain the context $c$ and the \emph{label} question $q_t$  from a human (contrast to $\hat{q}_t$ in equation \ref{eq:problem} which denotes the \emph{generated} question). We append an additional field to the dataset, \emph{Topic} $t$, and create three novel datasets, 1) SQuAD+, 2) MixSQuAD, and 3) MixSQuAD2X for the T-CQG task. The process of generating the three datasets is presented in figure \ref{fig:datasets}.

\subsubsection{Linking the target topic to data points, SQuAD+ dataset} \label{sec:squad+}
To identify semantic annotations for every context and question, we employ wikification \cite{WAT2014}, which annotates text inputs with relevant concepts from Wikipedia ($T_c$). We retain the top 5 concepts for each text (context and question) based on their PageRank scores, which reflect the authority of the concept over the annotated text. To make sure that we can link the topical alignment between the question and the context, we only retain examples where at least one common Wikipedia concept is present between the context and the question pair (i.e. |$T_c \cap T_q| \geq 1$). We select the concept with the highest PageRank score in the question (most authoritative) as the target topic $t$. This method ensures that the most closely related annotation is selected as the topic for each pair, and confirms that the topic is appropriately aligned with both the context and the question, thus avoiding situations where the topic may be relevant to one but not the other. As a result, both datasets have been enhanced to include paragraph-level contexts $c$, identified topics $t$, and corresponding questions $q_t$, as shown in figure \ref{fig:datasets}.


\subsubsection{MixSQuAD dataset} \label{sec:mixsquad}

We also create an enhanced dataset to synthesise a contrastive learning setting while fine-tuning the PLM for T-CQG leading to the \emph{MixSQuAD} dataset. When creating this dataset, we randomly pick pairs of observations from the SQuAD+ dataset described in section \ref{sec:squad+}. For each pair of examples $i$ and $j$ containing $(c_i, t_i, q_{t_i})$ and $(c_j, t_j, q_{t_j})$ respectively, we create two new examples where they share a common context $c_ic_j$ where the two contexts are concatenated. The data representation of the MixSQuAD dataset is presented in figure \ref{fig:datasets}. This approach aims to enhance the model's understanding of topics and the relationship between context, topic, and question by serving novel contrastive examples. An added benefit of the novel MixSQuAD dataset is that the context presented to the model during fine-tuning is guaranteed not to be previously encountered in the large corpora used for training foundational models. This method results in a diverse collection of 10,000 mixed data entries in the MixSQuAD dataset, fostering a robust learning environment for the models.

\subsubsection{MixSQuAD2X dataset} \label{sec:mixsquad2x}

The MixSQuAD2X dataset is very similar to MixSQuAD dataset, but the main difference is the utilisation of data augmentation to expand the dataset. In contrast to MixSQuAD, we introduce two additional examples to the dataset with the context $c_2c_1$ by reversing the order when concatenating the two randomly chosen contexts. This leads to a dataset that is twice as big as the MixSQuAD dataset. 


\subsection{Developing T-CQG Models for Education} \label{sec:models}

With the relevant datasets created, we built multiple models to be evaluated in a series of experiments to answer the research questions outlined in section \ref{sec:problem}. All the models used in experiments are created by finetuning the \texttt{T5-Small} \cite{Raffel2022} model, a small Language Model (sLM) that has also been used for educational question generation in the past \cite{bulathwela2023scalable,fawzi2024humanlike}. We fine-tuned the foundational model (\texttt{t5-small} from HuggingFace library\footnote{\url{https://huggingface.co/google-t5/t5-small}}) using the Adam optimizer with a batch size of 64, the learning rate of \(1e-3\), and epsilon of \(1e-8\). We use a maximum sequence length of 512 for the encoder, and 128 for the decoder. We train all models for a maximum of 50 epochs with an early stopping based on the validation loss \footnote{\url{https://github.com/Cathgy/Topic-controllable-Question-Generator.git}}.

\subsubsection{Baseline Model to Answer RQ2} We conducted fine-tuning for T-CQG using the same finetuning approach used by \cite{martin2020} for controlling complexity in simplifying texts. We used the proposed SQuAD+ dataset (described in section \ref{sec:squad+}) to finetune the T5 PLM. 

\subsubsection{TopicQG to Answer RQ2} The key difference between the baseline model and the proposed TopicQG model lies in the data used for fine-tuning the T5-small model. We introduced the TopicQG model to contrastive examples using the novel dataset created, MixSQuAD (described in section \ref{sec:mixsquad}). Such mixed contexts, which may feature sentences with vastly differing concepts, are designed to enhance the T5 model's understanding of the semantic relationships between context $c$, topic $t$, and question $q_t$.

\subsubsection{TopicQGedu to Answer RQ3}\label{topicQedu} Further refining the approach, we developed TopicQGedu, which incorporates an additional pre-training step. In this approach, the sLM model undergoes further training with scientific text documents before being fine-tuned. This step is intended to imbue the model with scientific terminology and concepts, crucial for crafting high-quality educational questions \cite{bulathwela2023scalable}.

\subsubsection{Quantised TopicQG Models to Answer RQ4} Quantisation allows reducing the memory footprint of neural models significantly to enhance their scalability. To evaluate the degree of loss due to quantising the trained models, we created the quantised versions of the TopicQG model. We used 8-bit quantisation utilising the \emph{LLM.int8} algorithm \cite{dettmers2022llmint8} and 4-bit precision employing the \emph{QLoRa} algorithm \cite{dettmers2023qlora} to create \textbf{TopicQG8bit} and \textbf{TopicQG4bit} models respectively.

\subsubsection{TopicQG2X to Answer RQ5} This model is trained similarly to the Topic QG model, but it exploits data augmentation by being finetuned on the newly proposed MixSQuAD2X dataset (described in section \ref{sec:mixsquad2x}). The MixSQuAD2X dataset  effectively doubles its size by changing the order of concatenation of contexts, introducing new examples to finetune the model with. This strategy has the potential to enhance the model’s robustness and generalisation abilities, improving the relevance and educational value of the generated questions to the given topics.

\subsubsection{Example Questions Generated with Models for the Experiments}

Table \ref{tab:examples} presents a random set of topic-controlled question generations based on the context text provided in five different subject areas (Computing, Economics, Chemistry, Art, and Biology).




\begin{table*}[]
\begin{center}
\caption{A Sample of Randomly Selected Generations from the TopicQG Model for Different Subject Domains}
\centerline{\includegraphics[width=\linewidth]{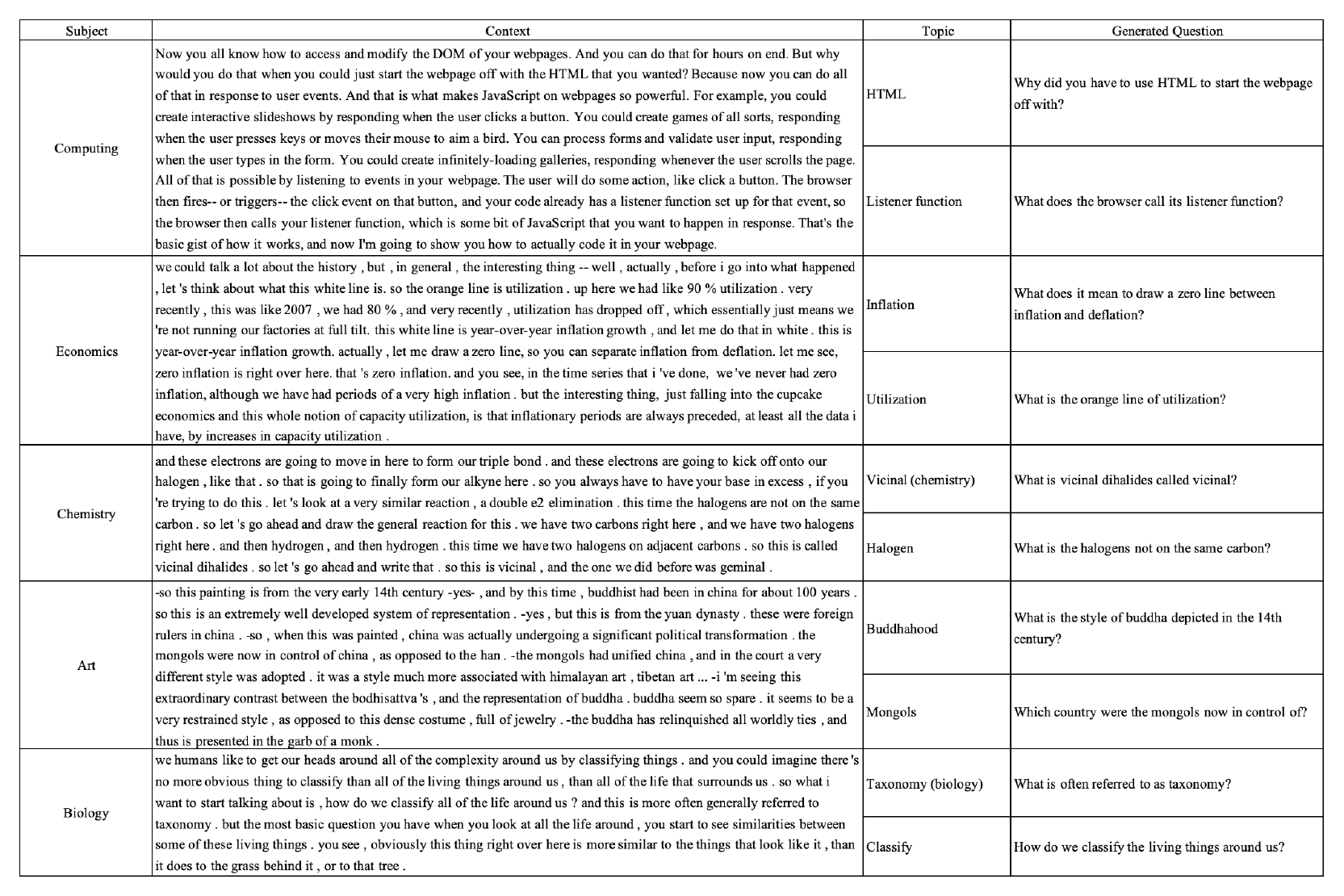}}
\label{tab:examples}
\end{center}
\end{table*}

\subsection{Human Annotation-based Evaluation of Semantic Relatedness Metrics}

To assess how representative BERTScore and WikiSemRel are when measuring topical relatedness (RQ1). We created a small gold-standard dataset via human annotation. The annotators (n = 4) consisted of two female and two male postgraduate students in the 20-30 age bracket from a masters degree programme at a university in the UK. 

To set up the annotation task, we randomly selected 30 questions from the MixKhanQ dataset (KhanQ dataset transformed using the method described in section \ref{sec:mixsquad}). For each sample, we provided the participants with the reference question $q_t$ and two corresponding generated questions, 1) the question  $\hat{q}_t$ generated using the \emph{relevant/prescribed} topic and 2) the question $\hat{q}_{t'}$ generated with an \emph{alternative} topic. Annotators were required to independently determine which of the two generated questions $\hat{q}_{t}$ or $\hat{q}_{t'}$ is more closely aligned with the reference question${q}_{t}$, the same tasks the SemRel metrics are going to do. The generated question $\hat{q}_{(\cdot)}$ annotators selected as closely relevant to the reference question is given $1$ and the other $0$. We calculated the Mean Absolute Error (MAE) between the mean score assigned by human annotators and the respective SemRel Score as per equation \ref{eq:mae}. 

\begin{equation} \label{eq:mae}
    \text{MAE(Human, SimRel)} = \frac{\sum_{\hat{q} \in Q} | \text{ Human}(\hat{q}) - \text{SemRel}(\hat{q}) |}{|Q|} \text{ where } Q \in \{q^{1}_{t}, q^{1}_{t'}, q^{2}_{t}, q^{2}_{t'}, \dots, q^{30}_{t}, q^{30}_{t'} \}
\end{equation}

\subsection{Experimental Setup for Automated Performance Evaluations}

\begin{figure*}[h]
\begin{center}
\centerline{\includegraphics[width=.7\linewidth]{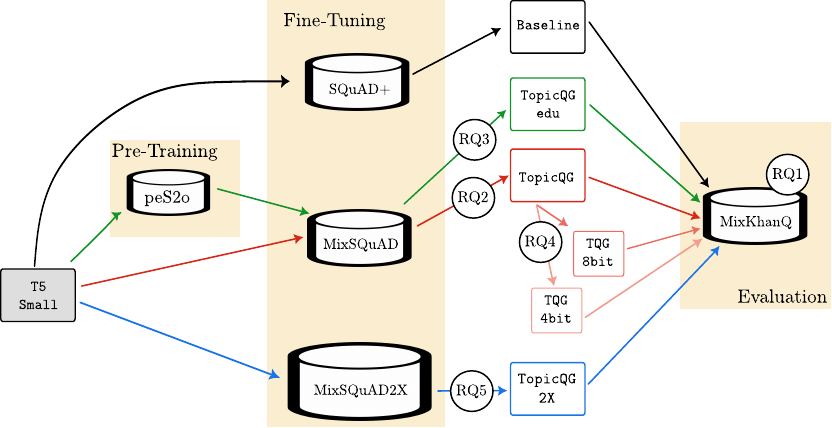}}
\caption{Methodology for training and evaluating the Baseline model (black),  TopicQGedu model (green, RQ3), TopicQG model (dark red, RQ2), its post-training quantised counterparts, TopicQG8bit model(medium red, RQ4), TopicQG4bit model(light red, RQ4) and TopicQG2X model (blue, RQ5). The numbered circles indicate different experimental pathways that test different research questions. The shaded grey box indicates that the \texttt{T5-Small} model was available pre-trained prior to the experiments while the non-shaded models contained parameters trained during the experiments.}
\label{fig:experiment}
\end{center}

\end{figure*}

Figure \ref{fig:experiment} illustrates the experimental setup designed to address RQs 2-5. 
A total of six models (including TopicQG's base, 8bit, and 4bit versions) have been developed as described in detail in section \ref{sec:models} and represented as coloured boxes in figure \ref{fig:experiment}. Each model is evaluated using the MixKhanQ dataset.

\subsection{Evaluation Metrics}

When evaluating the final models, we focused on two main aspects. 1) The generated question $\hat{q_t}$ is of high linguistic quality so it has the potential to be used in educational settings, 2) The generated question $\hat{q_t}$ is \emph{semantically related} to the prescribed topic $t$ so that it can address the AI-generated questions' common problem of being "too general to be useful in practice" in educational settings. 

\subsubsection{Evaluating the quality of generations} 

To assess the quality of the generated questions, the similarity between the reference question and the generated question is measured. We employed a suite of metrics, including BLEU \cite{papineni2002bleu}, METEOR\cite{banerjee2005meteor}, ROUGE \cite{lin2004rouge}, F1 score, and Perplexity\cite{HansenPPL}, which have been used frequently in previous research \cite{bulathwela2023scalable}. These metrics provide a comprehensive evaluation of the fluency, relevance, and coherence of the generated questions, serving as scalable indicators of the automated evaluation of the generated questions' quality.

\subsubsection{Semantic relatedness between the questions generated and the topic} 
For measuring the semantic relatedness, $SemRel(q_t, \hat{q}_t)$, we needed metrics that can quantify the relatedness between the reference question $q_t$ and the generated question $\hat{q}_t$. We used the \texttt{BERTScore} \cite{zhang2020bertscore} and the Wikipedia-based Topic Semantic Relatedness (\texttt{WikiSemRel}) \cite{Tagme} metrics for these evaluations. 

 \paragraph{\textbf{BERTScore}} leverages BERT contextual embeddings of tokens to calculate the similarity between two text extracts, improving upon the traditional exact match methods. Our early experiments showed that the BERTSCore tend to inflate the similarity between $q_t$ and $\hat{q}_t$, as there are words like "what" and "why" that overlap even if the generated question is not about the salient topic $t$ of the reference question. Therefore, we excluded stopwords in the reference and generated questions prior to calculating the BERTScore. BERTSCore is a score in the range (0,1) where 0 indicates no relatedness.

\paragraph{\textbf{WikiSemRel}} quantifies the semantic relatedness between the Wikipedia-based concepts extracted from the reference question $q_t$ and the generated question $\hat{q}_t$. We employ the WAT API \cite{WAT2014} service to calculate semantic relatedness using the 1) w2v-based method, that builds embeddings for Wiki entities based on their co-occurrence in Wikipedia pages and 2) Jaccard-based measure, that uses the outward links to other Wikipedia pages to calculate similarity \cite{Ponza2020}. We Wikify the generated question to compute the WikiSemRel score which is within range (0,1) where 0 indicates no relatedness. 



\section{Results}

In this section, we present the results from the experiments described in section \ref{methodology}. The results of the human evaluations answering RQ1 is presented in table \ref{tab:mae}. The offline evaluations to validate RQ 2-5 following the methodology illustrated in figure \ref{fig:experiment} are summarised in tables \ref{tab:QG} and \ref{tab:topic}. While table \ref{tab:QG} presents metrics relating to the linguistic quality of the generation, Table \ref{tab:topic} presents the semantic closeness between the prescribed topic and the generated questions. The perplexity calculation in table \ref{tab:QG} is done using the \texttt{TextDescriptives} python library with the \texttt{en\_core\_web\_lg} language model as the reference language distribution \cite{HansenPPL}.



\begin{table}[H]
\caption{Alignment between human annotation and Semantic Relatedness (SemRel) scores. The best performance and the next best for each metric is highlighted in \textbf{bold} and \textit{italic}.} \label{tab:mae}
\begin{tabular}{ c c c c }
\hline
    & BERT & WikiSemRel & WikiSemRel  \\ 
   &Score & (w2v) &  (Jaccard)
    \\\hline
$\text{MAE(Human, SemRel)} \downarrow$ & 0.48      & \textit{0.36}            & \textbf{0.23}    \\ \hline             
\end{tabular}
\end{table}


\begin{table*}[h]
\caption{Evaluation metrics relating to the quality of the questions generated by the models proposed in section \ref{sec:models} based on the MixKhanQ dataset. The best performance and the next best for each metric is highlighted in \textbf{bold} and \textit{italic}.}\label{tab:QG}
\begin{tabular}{l c c c c c c c c}
\hline
{Model/Metric} & {BLEU1$\uparrow$} & {BLEU2$\uparrow$} & {BLEU3$\uparrow$} &{BLEU4$\uparrow$} & {F1 Score$\uparrow$} & {METEOR$\uparrow$} & {Perplexity$\downarrow$} & {ROUGE-L$\uparrow$} \\ \hline
Baseline             & 0.519                     & 0.316                     & 0.216                     & 0.175                     & 0.319                        & 0.216                      & \textbf{1.303}                 & 0.207                       \\
TopicQGedu           & \textbf{0.551}            & 0.335                     & 0.221                     & 0.177                     & 0.302                        & 0.216                      & 1.360                          & 0.204                       \\
TopicQG              & \textbf{0.551}            & \textbf{0.343}            & \textbf{0.236}            & \textbf{0.191}            & \textbf{0.330}               & \textbf{0.233}             & \textit{1.323}                 & \textbf{0.230}              \\
\multicolumn{1}{r}{8bit}          & \textit{0.546}            & \textit{0.339}            & \textit{0.231}            & \textit{0.186}            & 0.319                        & \textit{0.226}             & 1.327                          & \textit{0.225}              \\
\multicolumn{1}{r}{4bit}         & 0.543                     & 0.337                     & \textit{0.231}            & \textit{0.186}            & 0.318                        & 0.223                      & 1.334                          & 0.223                       \\
TopicQG2X            & 0.536                     & 0.328                     & 0.221                     & 0.177                     & \textit{0.321}               & 0.220                      & 1.345                          & 0.216 \\ \hline                      
\end{tabular}

\end{table*}


\begin{table*}[]
\caption{Semantic relatedness between the generated questions $\hat{q}$ on (i) prescribed topic ${t}$ vs. (i) alternative topic ${t'}$ and the reference question on the prescribed topic ${q}_{t}$. The best performance and the next best for each metric is highlighted in \textbf{bold} and \textit{italic}.}
\label{tab:topic}
\begin{tabular}{l c c c c c c} \hline
                         & \multicolumn{3}{c}{BERTScore} & \multicolumn{3}{c}{WikiSimRel (Jaccard)} \\
                         &  $\hat{q}_t \uparrow$ & $\hat{q}_{t'} \downarrow$ & Difference $\uparrow$ &$ \hat{q}_t \uparrow$ & $\hat{q}_{t'} \downarrow$ & Difference $\uparrow$ \\ \hline
Baseline                 & \textbf{0.859}   & {0.859}            & {0.000}         & {0.615}          & \textit{\textbf{0.070}} & 0.545          \\
TopicQGedu               & {0.855}          & 0.831              & 0.024                  & 0.721            & 0.185                  & 0.536          \\
TopicQG                  & \textbf{0.859}   & \textit{0.830}     & \textit{0.029}         & \textit{0.727}   & \textit{0.132}         & \textit{0.595} \\
\multicolumn{1}{r}{8bit} & \textit{0.858}   & 0.831              & \textit{0.027}         & 0.693            & 0.142                  & 0.551          \\
\multicolumn{1}{r}{4bit}  & \textit{0.858}   & 0.831              & 0.027                  & 0.686            & 0.157                  & 0.529          \\
TopicQG2X                & \textbf{0.859}   & \textbf{0.823}     & \textbf{0.036}         & \textbf{0.735}   & \textbf{0.055}         & \textbf{0.680} \\ \hline
\end{tabular}
\end{table*}

\subsection{Most Representative Automated Topic Relevance Metric to Human Evaluations (RQ1)}

In human evaluations of the 30 randomly selected question pairs for topical alignment, only two pairs did not reach a consensus among the participants with one outlier in each case (with the
Fleiss' kappa \cite{fleiss1971measuring} measure of inter-rater agreement among multiple raters being 0.933). This indicates strong inter-rater reliability in the gold-standard human evaluator data. As per table \ref{tab:mae}, the WikiSimRel metrics are more aligned with the human judgements in comparison to the BERTScore. Among the three candidates, we can observe the embedding based (BERT and w2v) methods showing inferior representativeness. This could be due to the fact that embeddings can represent many different attributes about the entities and tokens they represent (e.g. whether the text is a question or a statement). This hypothesis is further reinforced by previous observations that including stopwords like "what", "why" leads to the inflation of BERTScore. However the Jaccard WikiSimRel score that relies exclusively on outward links from Wikipedia pages tends to capture a better representation of the informational and thematic content leading to better alignment. 



\subsection{Topical Relevance and the Effect of Pre-training on Generated Questions (RQ 2 and RQ 3)}

Table \ref{tab:QG} provides us an indication of the degree to which the generated question $\hat{q}_t$ resembles the reference question $q_t$. This is a proxy for topical relevance as the reference question is implicitly aligned with the controlled topic. The results indicate that the proposed TopicQG model outperforms the baseline model in all but perplexity metric. Outperforming in terms of BLEU scores at multiple levels (BLEU1 through BLEU4), indicates enhanced linguistic precision in question generation. It also achieves higher F1, ROUGE-L, and METEOR scores, reflecting the model's capability to generate questions that are not only relevant and accurate but also semantically aligned with reference texts. Compared with the baseline, a slight increase in perplexity suggests that the TopicQG model may generate questions that diverge from the reference language, potentially due to its ability to learn more complex educational expressions. The perplexity does not raise significant concerns over the quality of generations as the random examples in table \ref{tab:examples} doesn't indicate visible signs of deterioration. 
{It is noteworthy that the randomly selected examples in table \ref{tab:examples} are not as good as typical questions generated using a very large language model such as ChatGPT. We hypothesise the size of our model being a main reason for the relatively low quality of generations. However, our own prior work has also shown that such generations can be improved to humanly acceptable levels by simply post-processing them through a pre-trained grammar correction model \cite{fawzi2024humanlike,fawzi2023small}} retaining the accessibility and sustainability benefits of sLMs.  

Table \ref{tab:topic}, the stronger indicator of topic alignment gives us evidence that the proposed TopicQG models significantly outperform the baseline. In terms of the semantic difference between the educational questions generated with the controlled topic vs. a different topic (using WikiSimRel (Jaccard), the most representative metric from table \ref{tab:mae}), all newly proposed models except the 4bit quantised TopicQG model outperforms the baseline. This can be expected as extreme quantisation can deteriorate the accuracy of the model.  

In terms of the TopicQGedu model that is pre-trained on scientific text, the results are mixed and more difficult to interpret. While it surpasses the predictive performance on the Baseline in a few metrics, it performs below the TopicQG model across all metrics in table \ref{tab:QG}. While pre-training on scientific content is hypothesised to improve the topical relevance of the model, we do not observe improvements in this case. To rigorously assess whether the observed differences in performance metrics are statistically significant, we alsoconducted a paired t-test comparing the performance scores of TopicQGedu and TopicQG across the same set of questions. The results yielded a p-value of 0.083 (>0.05), suggesting that there is no statistically significant difference that TopicQGedu underperforms compared to TopicQG. Given that the T5 model is primarily trained on web-crawled data and Wikipedia articles \cite{Raffel2022}, the absence of scientific texts in the training corpus could potentially weaken the model’s performance in scientific concepts and language. Thus pre-training strategies may need to be further explored, especially in specialized domains where deeper domain knowledge might be crucial, even if immediate improvements in conventional metrics are not evident.

\subsection{Impact of Model Quantisation (RQ4) and Data Augmentation (RQ5)}

We investigated the effects of 8-bit and 4-bit quantisation on the TopicQG model (the best-performing model on the MixSQuAD dataset), referred to as TopicQG8bit and TopicQG4bit respectively. In comparison to the TopicQG model, the quantised models retain best performance with respect to metrics such as BLEU, F1-Score, MATEOR and ROUGE-L with very minor decreases ($\leq 0.01$) according to table \ref{tab:QG}. As expected, a drop in performance in comparison to the TopicQG model (with no quantisation) is observed. Similarly in table \ref{tab:topic}), a small drop in metrics is observed although it is not a drastic difference. This can be attributed to the fact that the generations change to a very small degree with quantisation indicated by the small deviations in table \ref{tab:QG}.  

Regarding memory usage, the full-precision TopicQG model occupies a memory size of $\approx 230$ MB. In contrast, the TopicQG8bit model significantly reduces this footprint to $\approx 110$ MB (59\%), and the TopicQG4bit model further reduces it to $\approx 94$ MB (53\%).  The potential of quantisation demonstrated in this study is twofold: 1) it significantly lowers the hardware requirements for running the models, and 2) it maintains a satisfactory level of performance, making it feasible to deploy educational topic-controllable question generation on platforms where computational resources are limited. This accessibility could dramatically widen the applications of such models, making them more ubiquitous in educational and other real-time interactive applications on mobile devices.  The reduction in model size not only implies lower memory requirements but also suggests lower power consumption, leading to cheaper infrastructure costs and a lower carbon footprint. Such properties are crucial for deploying these models in educational contexts of resource-constrained environments such as middle and low-income countries, mobile devices and embedded systems.

The comparisons between TopicQG and TopicQG2X  models in table \ref{tab:topic} show that the data augmentation has an obvious effect on improving the models performance on topical relevance. The greater diversity of examples where the same example is presented to the model in two different ways helps the model better understand to follow the topical theme prescribed in the instruction with the context. It surpasses all other models, including TopicQG, demonstrating superior alignment of the generated questions with the input context and topic. This highlights the effectiveness of data augmentation in enhancing the model’s capacity to generate questions with topic relevance and better contextual consideration of texts. 


\section{Discussion}

This paper tackled the challenge of topic-based educational question generation with a high degree of specificity. Due to the novelty of the task itself, we evolved our method over multiple steps to propose a method that can lead to high-quality T-CQG while validating novel approaches to evaluate the topical relevance of such generations. The results show that the novel method proposed and evaluated here is capable of generating topical educational questions while retaining coherent grammatical structure. Further experiments also showed how data augmentation increases the model's performance in topical relevance leading to improved results. The final experiments exploring quantisation indicate that the model's memory footprint can be halved with minimal loss of generative performance. Supported by human evaluation, the findings provide solid evidence that the questions generated by the proposed model are of high quality and meaningfully related to the educational content and topics, thereby affirming the effectiveness of our topic-controllable educational question generator.

Similar to trends in educational research in general  \cite{denny2024generativeaiedu}, the interest in the use of Generative Artificial Intelligence (GenAI) in LA research community has significantly increased in recent years. Regarding content generation, LLMs are used in tackling challenges such as grammar/ code correction \cite{cotet2020grammatical,viet2023Bug}, question generation \cite{elkins2024teachers,fawzi2024humanlike}, explanations and hints provision \cite{pardos2023hints,li2024Feedback}, in STEM subjects such as mathematics \cite{amini2019math,cobbe2021math} and science \cite{malinka2023Science,elkins2024teachers,bulathwela2023scalable}  to non-STEM domains  like law \cite{cui2023chatlaw} and language learning \cite{caines2023application}. Nevertheless, the majority of the community resorts to in-context learning \cite{dong2022survey} within enormous LLMs such as ChatGPT \cite{denny2024generativeaiedu}. For instance, there are increasing numbers of attempts of topic-controlled EdQG relying on Model-as-a-Service (MaaS) products that use externally hosted LLMs like ChatGPT (e.g. \cite{elkins2024teachers}). While practically valuable to varying degrees of success, these approaches introduce significant privacy, ethics, and governance challenges \cite{giannakos2024promise}. The extensive costs associated with the training and deployment of these models on-premise also make them impractical for educational stakeholders from both operational and financial perspectives \cite{fawzi2024humanlike}. 

We argue that over-reliance on such commercial models in academic research is a threat to academic independence and encourages alternative investigations to address significant challenges of education. The novel approach proposed in this paper provides significant opportunities to enhance the applicability of language models in educational contexts for question generation without the limitations posed by approaches relying on externally hosted LLMs like ChatGPT. The model proposed here has the potential to be scaled at a minimal cost in a safe and ethical manner and can be utilised to generate questions that are closely aligned with the specific content of educational materials. The 4-bit quantisation described reduces the model size to 94.41 MB while preserving essential performance, showcasing its potential for widespread use in resource-limited educational scenarios such as mobile devices and embedded systems. Therefore, the model has the potential for decreasing teachers workload on question generation in diverse contexts as well as being utilised in LMSs and ITSs to facilitate personalised learning experiences, allowing educational questions posed to be customised to meet the unique needs and interests of each learner (such as a learner model \cite{qiu2024toolbox}). 



\subsection{Implications of the Results for Research and Practice}

Regarding educational practice, the proposed topic-controlled question generating model can be useful for different tasks within the education domain. Primarily, we see such a tool as a teacher assistant tool to propose questions to teachers to select from. Such a tool would keep teachers in the loop as final decision makers but help them with tasks such as generating topic-specific, relevant, and age-appropriate questions for teaching. As discussed in the introduction, these tasks have long been identified as time-intensive tasks for teachers \cite{giannakos2024promise}. We envision tools where a teacher can point the system to a video, a presentation or a collection of learning resources where the system will automatically detect numerous salient topics and present them to the teacher as potential educational questions and the draft of a new quiz can be created in a matter of few clicks. This approach has the potential to change the degree of formative assessment due to decreased workload and can further stimulate well-anticipated innovation in education systems \cite{luckin2019designing}. We argue that the model proposed and evaluated here has the potential to decrease teachers' workload on such tasks. 

Second, the model can be integrated into multiple roles within the learning analytics infrastructures. The key to a precise learner state representation is having precise tests that can verify skill mastery of individuals at finer grain. The proposed method can lead to tools that can generate high-precision assessments within a personalised learning management system that can feed better data into learning analytics. While investing significant resources to create a relatively high coverage question banks is still feasible for short course and MOOC platforms that focus on narrow scopes of knowledge, as the world is gradually moving towards informal, lifelong learning such an investment would be infeasible. Models such as the one proposed here can play a critical role for continuous topic-specific, high quality and relevant question generation in educational systems. 

Third, the model can also bring efficiencies to the implementation of question generation in mobile and resource scarce contexts. As we are dealing with very small models, systems built on these models are scalable with minimal costs and has the potential to run on mobile devices without having to connect to the Internet. These considerations are of utmost importance for more equitable use of AI in Education \cite{bulathwela2024artificial}. 

Regarding LA and AI in Education research in general, the methodology proposed in this work can be extended to other forms of generation such as feedback, explanations, and content summaries in education. Also, aspects that that go beyond topical relevance (such as linguistic complexity and cognitive load etc.) can be controlled in future explorations to further advance learning analytics systems paving the way to re-imagining the limits of personalised learning material generation with AI. Furthermore, existing research either ignores the evaluation of whether the generated questions truly respond to the controlled conditions, or relies on extensive manual scoring by humans, which is both time-consuming and labour-intensive \cite{Yudelson13}. Experimental studies with human participants presented here indicate that the "relatedness score" has the potential to serve as a robust evaluation metric for assessing the semantic relatedness of generated questions to the input text, particularly in educational question generation tasks. It appears to excel in distinguishing between different concepts within the same academic field, making it particularly relevant for educational question generation tasks. As educational content generation research increases in LA literature, the importance of evaluation metrics of such content becomes even more important and the findings of this paper can help researchers consider appropriate metrics.



\section{Conclusion}

This paper proposes a novel approach to fine-tuning pre-trained sLMs to effectively address the challenge of generating topic-controllable questions based on paragraph-level context within educational settings. In addition, a novel method to synthesise training data for this task is presented with a novel Wikipedia concept-based evaluation method. The results show that the model proposed here has the potential to decrease teacher workload and improve personalised learning platforms also proving the effectiveness of training data. The model can also be scaled financially and operationally at a minimal cost to decrease academic researchers' over-reliance on commercial LLMs like ChatGPT. 

This study, while advancing topic-controllable question generation in education, acknowledges several limitations.  The limited human evaluation sample size hinders the statistical power of our findings about the semantic relatedness metrics although the extremely high inter-annotator agreement improves reliability of the result. More extensive human annotations would strengthen the results further. While we demonstrate the proposed novel method that randomly pairs contexts enabling the model T-CQG performance to improve, different pairing strategies that respect the subject domain, subtopics, difficulty level etc. can also lead to more effective training sets and should be explored in future studies. Finally, while the proposed method can be used to train the pre-trained model to contextualise generations to topical relevance, it focuses on topical relevance only. However, how to incorporate multiple aspects in addition to topical relevance such as linguistic complexity and generation length (e.g. short question) together should be explored in the future.  

\begin{acks}
This work is funded by the European Commission-funded projects "Humane AI" (Grant No. 820437) and "X5GON" (Grant No. 761758). This research is also part of the Teacher-AI Complementarity (TaiCo) project funded by the European Commission's Horizon Program (Project ID: 101177268).
\end{acks}

\bibliographystyle{ACM-Reference-Format}
\bibliography{sample-base}


\begin{thebibliography}{60}


\ifx \showCODEN    \undefined \def \showCODEN     #1{\unskip}     \fi
\ifx \showDOI      \undefined \def \showDOI       #1{#1}\fi
\ifx \showISBNx    \undefined \def \showISBNx     #1{\unskip}     \fi
\ifx \showISBNxiii \undefined \def \showISBNxiii  #1{\unskip}     \fi
\ifx \showISSN     \undefined \def \showISSN      #1{\unskip}     \fi
\ifx \showLCCN     \undefined \def \showLCCN      #1{\unskip}     \fi
\ifx \shownote     \undefined \def \shownote      #1{#1}          \fi
\ifx \showarticletitle \undefined \def \showarticletitle #1{#1}   \fi
\ifx \showURL      \undefined \def \showURL       {\relax}        \fi
\providecommand\bibfield[2]{#2}
\providecommand\bibinfo[2]{#2}
\providecommand\natexlab[1]{#1}
\providecommand\showeprint[2][]{arXiv:#2}

\bibitem[Adamson et~al\mbox{.}(2013)]%
        {Adamson2013}
\bibfield{author}{\bibinfo{person}{Derek Adamson}, \bibinfo{person}{Deepak
  Bhartiya}, \bibinfo{person}{Baljeet Gujral}, \bibinfo{person}{Ritu Kedia},
  \bibinfo{person}{Ankit Singh}, {and} \bibinfo{person}{Carolyn~P. Rose}.}
  \bibinfo{year}{2013}\natexlab{}.
\newblock \showarticletitle{Automatically Generating Discussion Questions}. In
  \bibinfo{booktitle}{\emph{Proceedings of the International Conference on
  Artificial Intelligence in Education (AIED)}}.
\newblock


\bibitem[Amini et~al\mbox{.}(2019)]%
        {amini2019math}
\bibfield{author}{\bibinfo{person}{Aida Amini}, \bibinfo{person}{Saadia
  Gabriel}, \bibinfo{person}{Peter Lin}, \bibinfo{person}{Rik
  Koncel-Kedziorski}, \bibinfo{person}{Yejin Choi}, {and}
  \bibinfo{person}{Hannaneh Hajishirzi}.} \bibinfo{year}{2019}\natexlab{}.
\newblock \showarticletitle{Mathqa: Towards interpretable math word problem
  solving with operation-based formalisms}.
\newblock \bibinfo{journal}{\emph{arXiv preprint arXiv:1905.13319}}
  (\bibinfo{year}{2019}).
\newblock


\bibitem[Bahrick et~al\mbox{.}(1993)]%
        {Bahrick1993}
\bibfield{author}{\bibinfo{person}{H.~P. Bahrick}, \bibinfo{person}{L.~E.
  Bahrick}, \bibinfo{person}{A.~S. Bahrick}, {and} \bibinfo{person}{P.~E.
  Bahrick}.} \bibinfo{year}{1993}\natexlab{}.
\newblock \showarticletitle{Maintenance of foreign language vocabulary and the
  spacing effect}.
\newblock \bibinfo{journal}{\emph{Psychological Science}} \bibinfo{volume}{4},
  \bibinfo{number}{5} (\bibinfo{year}{1993}), \bibinfo{pages}{316--321}.
\newblock


\bibitem[Banerjee and Lavie(2005)]%
        {banerjee2005meteor}
\bibfield{author}{\bibinfo{person}{S. Banerjee} {and} \bibinfo{person}{A.
  Lavie}.} \bibinfo{year}{2005}\natexlab{}.
\newblock \showarticletitle{METEOR: An automatic metric for MT evaluation with
  improved correlation with human judgments}. In \bibinfo{booktitle}{\emph{ACL
  Workshop on Intrinsic and Extrinsic Evaluation Measures for Machine
  Translation and/or Summarization}}. \bibinfo{pages}{65--72}.
\newblock


\bibitem[Blobstein et~al\mbox{.}(2023)]%
        {blobsteinangel}
\bibfield{author}{\bibinfo{person}{Ariel Blobstein}, \bibinfo{person}{Daniel
  Izmaylov}, \bibinfo{person}{Tal Yifat}, \bibinfo{person}{Michal Levy}, {and}
  \bibinfo{person}{Avi Segal}.} \bibinfo{year}{2023}\natexlab{}.
\newblock \showarticletitle{Angel: A New Generation Tool for Learning Material
  based Questions and Answers}. In \bibinfo{booktitle}{\emph{Proc.~of the
  NeurIPS Workshop on Generative AI for Education (GAIED)}}.
\newblock


\bibitem[Brank et~al\mbox{.}(2017)]%
        {Brank2017}
\bibfield{author}{\bibinfo{person}{Janez Brank}, \bibinfo{person}{Gregor
  Leban}, {and} \bibinfo{person}{Marko Grobelnik}.}
  \bibinfo{year}{2017}\natexlab{}.
\newblock \showarticletitle{Annotating Documents with Relevant Wikipedia
  Concepts}. In \bibinfo{booktitle}{\emph{Proc. of Slovenian KDD Conference on
  Data Mining and Data Warehouses (SiKDD)}}.
\newblock


\bibitem[Bulathwela et~al\mbox{.}(2023)]%
        {bulathwela2023scalable}
\bibfield{author}{\bibinfo{person}{Sahan Bulathwela}, \bibinfo{person}{Hamze
  Muse}, {and} \bibinfo{person}{Emine Yilmaz}.}
  \bibinfo{year}{2023}\natexlab{}.
\newblock \showarticletitle{Scalable educational question generation with
  pre-trained language models}. In \bibinfo{booktitle}{\emph{International
  Conference on Artificial Intelligence in Education}}. Springer,
  \bibinfo{pages}{327--339}.
\newblock


\bibitem[Bulathwela et~al\mbox{.}(2024)]%
        {bulathwela2024artificial}
\bibfield{author}{\bibinfo{person}{Sahan Bulathwela},
  \bibinfo{person}{Mar{\'\i}a P{\'e}rez-Ortiz}, \bibinfo{person}{Catherine
  Holloway}, \bibinfo{person}{Mutlu Cukurova}, {and} \bibinfo{person}{John
  Shawe-Taylor}.} \bibinfo{year}{2024}\natexlab{}.
\newblock \showarticletitle{Artificial intelligence alone will not democratise
  education: On educational inequality, techno-solutionism and inclusive
  tools}.
\newblock \bibinfo{journal}{\emph{Sustainability}} \bibinfo{volume}{16},
  \bibinfo{number}{2} (\bibinfo{year}{2024}), \bibinfo{pages}{781}.
\newblock


\bibitem[Bulathwela et~al\mbox{.}(2021)]%
        {bulathwela2021semantic}
\bibfield{author}{\bibinfo{person}{Sahan Bulathwela},
  \bibinfo{person}{Mar{\'\i}a P{\'e}rez-Ortiz}, \bibinfo{person}{Emine Yilmaz},
  {and} \bibinfo{person}{John Shawe-Taylor}.} \bibinfo{year}{2021}\natexlab{}.
\newblock \showarticletitle{Semantic TrueLearn: using semantic knowledge graphs
  in recommendation systems}.
\newblock \bibinfo{journal}{\emph{arXiv preprint arXiv:2112.04368}}
  (\bibinfo{year}{2021}).
\newblock


\bibitem[Caines et~al\mbox{.}(2023)]%
        {caines2023application}
\bibfield{author}{\bibinfo{person}{Andrew Caines}, \bibinfo{person}{Luca
  Benedetto}, \bibinfo{person}{Shiva Taslimipoor}, \bibinfo{person}{Christopher
  Davis}, {et~al\mbox{.}}} \bibinfo{year}{2023}\natexlab{}.
\newblock \showarticletitle{On the application of large language models for
  language teaching and assessment technology}.
\newblock \bibinfo{journal}{\emph{arXiv preprint arXiv:2307.08393}}
  (\bibinfo{year}{2023}).
\newblock


\bibitem[Chen et~al\mbox{.}(2021)]%
        {Chen2021}
\bibfield{author}{\bibinfo{person}{Feng Chen}, \bibinfo{person}{Jiayuan Xie},
  \bibinfo{person}{Yi Cai}, \bibinfo{person}{Tao Wang}, {and}
  \bibinfo{person}{Qing Li}.} \bibinfo{year}{2021}\natexlab{}.
\newblock \showarticletitle{Difficulty-Controllable Visual Question
  Generation}. In \bibinfo{booktitle}{\emph{Proc. Web and Big Data:
  International Joint Conference}}. \bibinfo{publisher}{Springer-Verlag},
  \bibinfo{pages}{332--347}.
\newblock


\bibitem[Cobbe et~al\mbox{.}(2021)]%
        {cobbe2021math}
\bibfield{author}{\bibinfo{person}{Karl Cobbe}, \bibinfo{person}{Vineet
  Kosaraju}, \bibinfo{person}{Mohammad Bavarian}, \bibinfo{person}{Mark Chen},
  \bibinfo{person}{Heewoo Jun}, \bibinfo{person}{Lukasz Kaiser},
  \bibinfo{person}{Matthias Plappert}, \bibinfo{person}{Jerry Tworek},
  \bibinfo{person}{Jacob Hilton}, \bibinfo{person}{Reiichiro Nakano},
  {et~al\mbox{.}}} \bibinfo{year}{2021}\natexlab{}.
\newblock \showarticletitle{Training verifiers to solve math word problems}.
\newblock \bibinfo{journal}{\emph{arXiv preprint arXiv:2110.14168}}
  (\bibinfo{year}{2021}).
\newblock


\bibitem[Cotet et~al\mbox{.}(2020)]%
        {cotet2020grammatical}
\bibfield{author}{\bibinfo{person}{Teodor-Mihai Cotet}, \bibinfo{person}{Stefan
  Ruseti}, {and} \bibinfo{person}{Mihai Dascalu}.}
  \bibinfo{year}{2020}\natexlab{}.
\newblock \showarticletitle{Neural grammatical error correction for romanian}.
  In \bibinfo{booktitle}{\emph{2020 IEEE 32nd International Conference on Tools
  with Artificial Intelligence (ICTAI)}}. IEEE, \bibinfo{pages}{625--631}.
\newblock


\bibitem[Cui et~al\mbox{.}(2023)]%
        {cui2023chatlaw}
\bibfield{author}{\bibinfo{person}{Jiaxi Cui}, \bibinfo{person}{Zongjian Li},
  \bibinfo{person}{Yang Yan}, \bibinfo{person}{Bohua Chen}, {and}
  \bibinfo{person}{Li Yuan}.} \bibinfo{year}{2023}\natexlab{}.
\newblock \showarticletitle{Chatlaw: Open-source legal large language model
  with integrated external knowledge bases}.
\newblock \bibinfo{journal}{\emph{arXiv preprint arXiv:2306.16092}}
  (\bibinfo{year}{2023}).
\newblock


\bibitem[Cukurova et~al\mbox{.}(2023)]%
        {cukurova2023adoption}
\bibfield{author}{\bibinfo{person}{Mutlu Cukurova}, \bibinfo{person}{Xin Miao},
  {and} \bibinfo{person}{Richard Brooker}.} \bibinfo{year}{2023}\natexlab{}.
\newblock \showarticletitle{Adoption of artificial intelligence in schools:
  unveiling factors influencing teachers’ engagement}. In
  \bibinfo{booktitle}{\emph{International conference on artificial intelligence
  in education}}. Springer, \bibinfo{pages}{151--163}.
\newblock


\bibitem[Dathathri et~al\mbox{.}(2020)]%
        {dathathri2020plug}
\bibfield{author}{\bibinfo{person}{Sumanth Dathathri}, \bibinfo{person}{Andrea
  Madotto}, \bibinfo{person}{Janice Lan}, \bibinfo{person}{Jane Hung},
  \bibinfo{person}{Eric Frank}, \bibinfo{person}{Piero Molino},
  \bibinfo{person}{Jason Yosinski}, {and} \bibinfo{person}{Rosanne Liu}.}
  \bibinfo{year}{2020}\natexlab{}.
\newblock \showarticletitle{Plug and Play Language Models: A Simple Approach to
  Controlled Text Generation}. In \bibinfo{booktitle}{\emph{International
  Conference on Learning Representations}}.
\newblock
\urldef\tempurl%
\url{https://openreview.net/forum?id=H1edEyBKDS}
\showURL{%
\tempurl}


\bibitem[Denny et~al\mbox{.}(2024)]%
        {denny2024generativeaiedu}
\bibfield{author}{\bibinfo{person}{Paul Denny}, \bibinfo{person}{Sumit
  Gulwani}, \bibinfo{person}{Neil~T. Heffernan}, \bibinfo{person}{Tanja
  Käser}, \bibinfo{person}{Steven Moore}, \bibinfo{person}{Anna~N. Rafferty},
  {and} \bibinfo{person}{Adish Singla}.} \bibinfo{year}{2024}\natexlab{}.
\newblock \bibinfo{title}{Generative AI for Education (GAIED): Advances,
  Opportunities, and Challenges}.
\newblock
\newblock
\showeprint[arxiv]{2402.01580}~[cs.CY]
\urldef\tempurl%
\url{https://arxiv.org/abs/2402.01580}
\showURL{%
\tempurl}


\bibitem[Dettmers et~al\mbox{.}(2022)]%
        {dettmers2022llmint8}
\bibfield{author}{\bibinfo{person}{Tim Dettmers}, \bibinfo{person}{Mike Lewis},
  \bibinfo{person}{Younes Belkada}, {and} \bibinfo{person}{Luke Zettlemoyer}.}
  \bibinfo{year}{2022}\natexlab{}.
\newblock \showarticletitle{LLM.int8(): 8-bit Matrix Multiplication for
  Transformers at Scale}.
\newblock \bibinfo{journal}{\emph{arXiv preprint arXiv:2208.07339}}
  (\bibinfo{year}{2022}).
\newblock


\bibitem[Dettmers et~al\mbox{.}(2023)]%
        {dettmers2023qlora}
\bibfield{author}{\bibinfo{person}{Tim Dettmers}, \bibinfo{person}{Artidoro
  Pagnoni}, \bibinfo{person}{Ari Holtzman}, {and} \bibinfo{person}{Luke
  Zettlemoyer}.} \bibinfo{year}{2023}\natexlab{}.
\newblock \showarticletitle{QLora: Efficient Fine-Tuning of Quantized LLMs}.
\newblock \bibinfo{journal}{\emph{arXiv preprint arXiv:2305.14314}}
  (\bibinfo{year}{2023}).
\newblock


\bibitem[Do~Viet and Markov(2023)]%
        {viet2023Bug}
\bibfield{author}{\bibinfo{person}{Tung Do~Viet} {and}
  \bibinfo{person}{Konstantin Markov}.} \bibinfo{year}{2023}\natexlab{}.
\newblock \showarticletitle{Using Large Language Models for Bug Localization
  and Fixing}. In \bibinfo{booktitle}{\emph{2023 12th International Conference
  on Awareness Science and Technology (iCAST)}}. IEEE,
  \bibinfo{pages}{192--197}.
\newblock


\bibitem[Dong et~al\mbox{.}(2022)]%
        {dong2022survey}
\bibfield{author}{\bibinfo{person}{Qingxiu Dong}, \bibinfo{person}{Lei Li},
  \bibinfo{person}{Damai Dai}, \bibinfo{person}{Ce Zheng},
  \bibinfo{person}{Zhiyong Wu}, \bibinfo{person}{Baobao Chang},
  \bibinfo{person}{Xu Sun}, \bibinfo{person}{Jingjing Xu}, {and}
  \bibinfo{person}{Zhifang Sui}.} \bibinfo{year}{2022}\natexlab{}.
\newblock \showarticletitle{A survey on in-context learning}.
\newblock \bibinfo{journal}{\emph{arXiv preprint arXiv:2301.00234}}
  (\bibinfo{year}{2022}).
\newblock


\bibitem[Du et~al\mbox{.}(2017)]%
        {Du2017}
\bibfield{author}{\bibinfo{person}{Xinya Du}, \bibinfo{person}{Junru Shao},
  {and} \bibinfo{person}{Claire Cardie}.} \bibinfo{year}{2017}\natexlab{}.
\newblock \showarticletitle{Learning to Ask: Neural Question Generation for
  Reading Comprehension}. In \bibinfo{booktitle}{\emph{Proc. Annual Meeting of
  the Association for Computational Linguistics}}. \bibinfo{pages}{1342--1352}.
\newblock


\bibitem[Elkins et~al\mbox{.}(2024)]%
        {elkins2024teachers}
\bibfield{author}{\bibinfo{person}{Sabina Elkins}, \bibinfo{person}{Ekaterina
  Kochmar}, \bibinfo{person}{Jackie~CK Cheung}, {and} \bibinfo{person}{Iulian
  Serban}.} \bibinfo{year}{2024}\natexlab{}.
\newblock \showarticletitle{How Teachers Can Use Large Language Models and
  Bloom’s Taxonomy to Create Educational Quizzes}. In
  \bibinfo{booktitle}{\emph{Proceedings of the AAAI Conference on Artificial
  Intelligence}}, Vol.~\bibinfo{volume}{38}. \bibinfo{pages}{23084--23091}.
\newblock


\bibitem[Faraby et~al\mbox{.}(2024)]%
        {AlFaraby2024}
\bibfield{author}{\bibinfo{person}{Said~Al Faraby}, \bibinfo{person}{Ade
  Romadhony}, {and} \bibinfo{person}{Adiwijaya}.}
  \bibinfo{year}{2024}\natexlab{}.
\newblock \showarticletitle{Analysis of LLMs for educational question
  classification and generation}.
\newblock \bibinfo{journal}{\emph{Computers and Education: Artificial
  Intelligence}}  \bibinfo{volume}{7} (\bibinfo{year}{2024}),
  \bibinfo{pages}{100298}.
\newblock
\urldef\tempurl%
\url{https://doi.org/10.1016/j.caeai.2024.100298}
\showDOI{\tempurl}


\bibitem[Fawzi et~al\mbox{.}({[n.\,d.]})]%
        {fawzi2023small}
\bibfield{author}{\bibinfo{person}{Fares Fawzi}, \bibinfo{person}{Sadie Amini},
  {and} \bibinfo{person}{Sahan Bulathwela}.}
  \bibinfo{year}{[n.\,d.]}\natexlab{}.
\newblock \showarticletitle{Small Generative Language Models for Educational
  Question Generation}. In \bibinfo{booktitle}{\emph{Proc.~of the NeurIPS
  Workshop on Generative AI for Education (GAIED)}}.
\newblock


\bibitem[Fawzi et~al\mbox{.}(2024)]%
        {fawzi2024humanlike}
\bibfield{author}{\bibinfo{person}{F. Fawzi}, \bibinfo{person}{S. Balan},
  \bibinfo{person}{M. Cukurova}, \bibinfo{person}{E. Yilmaz}, {and}
  \bibinfo{person}{S. Bulathwela}.} \bibinfo{year}{2024}\natexlab{}.
\newblock \showarticletitle{Towards Human-Like Educational Question Generation
  with Small Language Models}. In \bibinfo{booktitle}{\emph{Artificial
  Intelligence in Education. Posters and Late Breaking Results, Workshops and
  Tutorials, Industry and Innovation Tracks, Practitioners, Doctoral Consortium
  and Blue Sky}}, Vol.~\bibinfo{volume}{2150}. \bibinfo{publisher}{Springer},
  \bibinfo{address}{Cham}.
\newblock


\bibitem[Ferragina and Scaiella(2010)]%
        {Tagme}
\bibfield{author}{\bibinfo{person}{Paolo Ferragina} {and} \bibinfo{person}{Ugo
  Scaiella}.} \bibinfo{year}{2010}\natexlab{}.
\newblock \showarticletitle{TAGME: on-the-fly annotation of short text
  fragments (by wikipedia entities)}. In \bibinfo{booktitle}{\emph{Proceedings
  of the 19th ACM International Conference on Information and Knowledge
  Management}} (Toronto, ON, Canada) \emph{(\bibinfo{series}{CIKM '10})}.
  \bibinfo{publisher}{Association for Computing Machinery},
  \bibinfo{address}{New York, NY, USA}, \bibinfo{pages}{1625–1628}.
\newblock
\showISBNx{9781450300995}
\urldef\tempurl%
\url{https://doi.org/10.1145/1871437.1871689}
\showDOI{\tempurl}


\bibitem[Fleiss(1971)]%
        {fleiss1971measuring}
\bibfield{author}{\bibinfo{person}{Joseph~L. Fleiss}.}
  \bibinfo{year}{1971}\natexlab{}.
\newblock \showarticletitle{Measuring nominal scale agreement among many
  raters}.
\newblock \bibinfo{journal}{\emph{Psychological Bulletin}}
  \bibinfo{volume}{76}, \bibinfo{number}{5} (\bibinfo{year}{1971}),
  \bibinfo{pages}{378--382}.
\newblock


\bibitem[for Education(2024)]%
        {gov2024use}
\bibfield{author}{\bibinfo{person}{Department for Education}.}
  \bibinfo{year}{2024}\natexlab{}.
\newblock \bibinfo{booktitle}{\emph{Use Cases for Generative AI in Education:
  User Research Report}}.
\newblock \bibinfo{type}{{T}echnical {R}eport}.
  \bibinfo{institution}{Department for Education, UK Government}.
\newblock
\urldef\tempurl%
\url{https://www.gov.uk/government/publications/generative-ai-in-education-user-research-and-technical-report}
\showURL{%
\tempurl}
\newblock
\shownote{Accessed: 2024-09-21}.


\bibitem[Giannakos et~al\mbox{.}(2024)]%
        {giannakos2024promise}
\bibfield{author}{\bibinfo{person}{Michail Giannakos}, \bibinfo{person}{Roger
  Azevedo}, \bibinfo{person}{Peter Brusilovsky}, \bibinfo{person}{Mutlu
  Cukurova}, \bibinfo{person}{Yannis Dimitriadis}, \bibinfo{person}{Davinia
  Hernandez-Leo}, \bibinfo{person}{Sanna J{\"a}rvel{\"a}},
  \bibinfo{person}{Manolis Mavrikis}, {and} \bibinfo{person}{Bart Rienties}.}
  \bibinfo{year}{2024}\natexlab{}.
\newblock \showarticletitle{The promise and challenges of generative AI in
  education}.
\newblock \bibinfo{journal}{\emph{Behaviour \& Information Technology}}
  (\bibinfo{year}{2024}), \bibinfo{pages}{1--27}.
\newblock


\bibitem[Gong and Pan(2022)]%
        {gong-etal-2022-khanq}
\bibfield{author}{\bibinfo{person}{Huanli Gong} {and}
  \bibinfo{person}{Hengchang Pan, Liangming~andHu}.}
  \bibinfo{year}{2022}\natexlab{}.
\newblock \showarticletitle{{KHANQ}: A Dataset for Generating Deep Questions in
  Education}. In \bibinfo{booktitle}{\emph{Proceedings of the 29th
  International Conference on Computational Linguistics}}.
\newblock


\bibitem[Hansen et~al\mbox{.}(2023)]%
        {HansenPPL}
\bibfield{author}{\bibinfo{person}{Lasse Hansen}, \bibinfo{person}{Ludvig~Renbo
  Olsen}, {and} \bibinfo{person}{Kenneth Enevoldsen}.}
  \bibinfo{year}{2023}\natexlab{}.
\newblock \showarticletitle{TextDescriptives: A Python package for calculating
  a large variety of metrics from text}.
\newblock \bibinfo{journal}{\emph{Journal of Open Source Software}}
  \bibinfo{volume}{8}, \bibinfo{number}{84} (\bibinfo{date}{April}
  \bibinfo{year}{2023}), \bibinfo{pages}{5153}.
\newblock
\showISSN{2475-9066}
\urldef\tempurl%
\url{https://doi.org/10.21105/joss.05153}
\showDOI{\tempurl}


\bibitem[Heilman and Smith(2010)]%
        {HeilmanSmith2010}
\bibfield{author}{\bibinfo{person}{Michael Heilman} {and}
  \bibinfo{person}{Noah~A. Smith}.} \bibinfo{year}{2010}\natexlab{}.
\newblock \showarticletitle{Good question! Statistical ranking for question
  generation}. In \bibinfo{booktitle}{\emph{Proceedings of the Human Language
  Technology Conference and the North American Chapter of the Association for
  Computational Linguistics (HLT-NAACL)}}.
\newblock


\bibitem[Hu et~al\mbox{.}(2018)]%
        {hu2018topic}
\bibfield{author}{\bibinfo{person}{Wenpeng Hu}, \bibinfo{person}{Bing Liu},
  \bibinfo{person}{Rui Yan}, \bibinfo{person}{Dongyan Zhao}, {and}
  \bibinfo{person}{Jinwen Ma}.} \bibinfo{year}{2018}\natexlab{}.
\newblock \showarticletitle{Topic-Based Question Generation}. In
  \bibinfo{booktitle}{\emph{International Conference on Learning
  Representations}}.
\newblock
\newblock
\shownote{ICLR 2018 Conference Blind Submission. Invite to Workshop Track}.


\bibitem[Khalifa et~al\mbox{.}(2021)]%
        {khalifa2021distributional}
\bibfield{author}{\bibinfo{person}{Muhammad Khalifa}, \bibinfo{person}{Hady
  Elsahar}, {and} \bibinfo{person}{Marc Dymetman}.}
  \bibinfo{year}{2021}\natexlab{}.
\newblock \showarticletitle{A Distributional Approach to Controlled Text
  Generation}. In \bibinfo{booktitle}{\emph{International Conference on
  Learning Representations}}.
\newblock
\urldef\tempurl%
\url{https://openreview.net/forum?id=jWkw45-9AbL}
\showURL{%
\tempurl}


\bibitem[Kuo et~al\mbox{.}(2023)]%
        {kuo2023knowledge}
\bibfield{author}{\bibinfo{person}{Bor-Chen Kuo}, \bibinfo{person}{Frederic~TY
  Chang}, {and} \bibinfo{person}{Zong-En Bai}.}
  \bibinfo{year}{2023}\natexlab{}.
\newblock \showarticletitle{Leveraging LLMs for Adaptive Testing and Learning
  in Taiwan Adaptive Learning Platform (TALP).}. In
  \bibinfo{booktitle}{\emph{Workshop on Empowering Education with LLMs at
  AIED}}. \bibinfo{pages}{101--110}.
\newblock


\bibitem[Li et~al\mbox{.}(2024)]%
        {li2024Feedback}
\bibfield{author}{\bibinfo{person}{Hai Li}, \bibinfo{person}{Chenglu Li},
  \bibinfo{person}{Wanli Xing}, \bibinfo{person}{Sami Baral}, {and}
  \bibinfo{person}{Neil Heffernan}.} \bibinfo{year}{2024}\natexlab{}.
\newblock \showarticletitle{Automated Feedback for Student Math Responses Based
  on Multi-Modality and Fine-Tuning}. In \bibinfo{booktitle}{\emph{Proceedings
  of the 14th Learning Analytics and Knowledge Conference}}.
  \bibinfo{pages}{763--770}.
\newblock


\bibitem[Lin(2004)]%
        {lin2004rouge}
\bibfield{author}{\bibinfo{person}{C.~Y. Lin}.}
  \bibinfo{year}{2004}\natexlab{}.
\newblock \showarticletitle{ROUGE: A package for automatic evaluation of
  summaries}. In \bibinfo{booktitle}{\emph{Workshop on Text Summarization
  Branches Out}}.
\newblock


\bibitem[Lopez et~al\mbox{.}(2021)]%
        {lopez2021simplifying}
\bibfield{author}{\bibinfo{person}{L.~E. Lopez}, \bibinfo{person}{D.~K. Cruz},
  \bibinfo{person}{J.~C.~B. Cruz}, {and} \bibinfo{person}{C. Cheng}.}
  \bibinfo{year}{2021}\natexlab{}.
\newblock \showarticletitle{Simplifying Paragraph-level Question Generation via
  Transformer Language Models}. In \bibinfo{booktitle}{\emph{Proceedings of the
  PRICAI 2021: Trends in Artificial Intelligence}} (8--12 November 2021).
  \bibinfo{address}{Hanoi, Vietnam}.
\newblock


\bibitem[Luckin and Cukurova(2019)]%
        {luckin2019designing}
\bibfield{author}{\bibinfo{person}{Rosemary Luckin} {and}
  \bibinfo{person}{Mutlu Cukurova}.} \bibinfo{year}{2019}\natexlab{}.
\newblock \showarticletitle{Designing educational technologies in the age of
  AI: A learning sciences-driven approach}.
\newblock \bibinfo{journal}{\emph{British Journal of Educational Technology}}
  \bibinfo{volume}{50}, \bibinfo{number}{6} (\bibinfo{year}{2019}),
  \bibinfo{pages}{2824--2838}.
\newblock


\bibitem[Malinka et~al\mbox{.}(2023)]%
        {malinka2023Science}
\bibfield{author}{\bibinfo{person}{Kamil Malinka}, \bibinfo{person}{Martin
  Peres{\'i}ni}, \bibinfo{person}{Anton Firc}, \bibinfo{person}{Ondrej
  Hujn{\'a}k}, {and} \bibinfo{person}{Filip Janus}.}
  \bibinfo{year}{2023}\natexlab{}.
\newblock \showarticletitle{On the educational impact of chatgpt: Is artificial
  intelligence ready to obtain a university degree?}. In
  \bibinfo{booktitle}{\emph{Proceedings of the 2023 Conference on Innovation
  and Technology in Computer Science Education}}, Vol.~\bibinfo{volume}{1}.
  \bibinfo{pages}{47--53}.
\newblock


\bibitem[Martin et~al\mbox{.}(2020)]%
        {martin2020}
\bibfield{author}{\bibinfo{person}{Louis Martin}, \bibinfo{person}{{\'E}ric
  Villemonte~de La~Clergerie}, \bibinfo{person}{Beno{\^i}t Sagot}, {and}
  \bibinfo{person}{Antoine Bordes}.} \bibinfo{year}{2020}\natexlab{}.
\newblock \showarticletitle{{Controllable Sentence Simplification}}. In
  \bibinfo{booktitle}{\emph{{LREC 2020 - 12th Language Resources and Evaluation
  Conference}}}. \bibinfo{address}{Marseille, France}.
\newblock
\urldef\tempurl%
\url{https://inria.hal.science/hal-02678214}
\showURL{%
\tempurl}


\bibitem[Papineni et~al\mbox{.}(2002)]%
        {papineni2002bleu}
\bibfield{author}{\bibinfo{person}{K. Papineni}, \bibinfo{person}{S. Roukos},
  \bibinfo{person}{T. Ward}, {and} \bibinfo{person}{W.J. Zhu}.}
  \bibinfo{year}{2002}\natexlab{}.
\newblock \showarticletitle{BLEU: a method for automatic evaluation of machine
  translation}. In \bibinfo{booktitle}{\emph{Proceedings of the 40th annual
  meeting on association for computational linguistics}}. Association for
  Computational Linguistics, \bibinfo{pages}{311--318}.
\newblock


\bibitem[Pardos and Bhandari(2023)]%
        {pardos2023hints}
\bibfield{author}{\bibinfo{person}{Zachary~A Pardos} {and}
  \bibinfo{person}{Shreya Bhandari}.} \bibinfo{year}{2023}\natexlab{}.
\newblock \showarticletitle{Learning gain differences between ChatGPT and human
  tutor generated algebra hints}.
\newblock \bibinfo{journal}{\emph{arXiv preprint arXiv:2302.06871}}
  (\bibinfo{year}{2023}).
\newblock


\bibitem[Piccinno and Ferragina(2014)]%
        {WAT2014}
\bibfield{author}{\bibinfo{person}{Francesco Piccinno} {and}
  \bibinfo{person}{Paolo Ferragina}.} \bibinfo{year}{2014}\natexlab{}.
\newblock \showarticletitle{From TagME to WAT: a new entity annotator}. In
  \bibinfo{booktitle}{\emph{Proceedings of the First International Workshop on
  Entity Recognition \& Disambiguation}} \emph{(\bibinfo{series}{ERD '14})}.
  \bibinfo{publisher}{Association for Computing Machinery},
  \bibinfo{pages}{55–62}.
\newblock
\showISBNx{9781450330237}
\urldef\tempurl%
\url{https://doi.org/10.1145/2633211.2634350}
\showDOI{\tempurl}


\bibitem[Pinto et~al\mbox{.}(2023)]%
        {pinto2023write}
\bibfield{author}{\bibinfo{person}{Gustavo Pinto}, \bibinfo{person}{Isadora
  Cardoso-Pereira}, \bibinfo{person}{Danilo Monteiro}, \bibinfo{person}{Danilo
  Lucena}, \bibinfo{person}{Alberto Souza}, {and} \bibinfo{person}{Kiev Gama}.}
  \bibinfo{year}{2023}\natexlab{}.
\newblock \showarticletitle{Large language models for education: Grading
  open-ended questions using chatgpt}. In \bibinfo{booktitle}{\emph{Proceedings
  of the XXXVII Brazilian Symposium on Software Engineering}}.
  \bibinfo{pages}{293--302}.
\newblock


\bibitem[Ponza et~al\mbox{.}(2020)]%
        {Ponza2020}
\bibfield{author}{\bibinfo{person}{Marco Ponza}, \bibinfo{person}{Paolo
  Ferragina}, {and} \bibinfo{person}{Soumen Chakrabarti}.}
  \bibinfo{year}{2020}\natexlab{}.
\newblock \showarticletitle{On Computing Entity Relatedness in Wikipedia, with
  Applications}.
\newblock \bibinfo{journal}{\emph{Knowledge-Based Systems}}
  \bibinfo{volume}{188} (\bibinfo{year}{2020}).
\newblock


\bibitem[Qiu et~al\mbox{.}(2024)]%
        {qiu2024toolbox}
\bibfield{author}{\bibinfo{person}{Yuxiang Qiu}, \bibinfo{person}{Karim
  Djemili}, \bibinfo{person}{Denis Elezi}, \bibinfo{person}{Aaneel~Shalman
  Srazali}, \bibinfo{person}{Mar{\'\i}a P{\'e}rez-Ortiz},
  \bibinfo{person}{Emine Yilmaz}, \bibinfo{person}{John Shawe-Taylor}, {and}
  \bibinfo{person}{Sahan Bulathwela}.} \bibinfo{year}{2024}\natexlab{}.
\newblock \showarticletitle{A Toolbox for Modelling Engagement with Educational
  Videos}. In \bibinfo{booktitle}{\emph{Proceedings of the AAAI Conference on
  Artificial Intelligence}}, Vol.~\bibinfo{volume}{38}.
  \bibinfo{pages}{23128--23136}.
\newblock


\bibitem[Raffel et~al\mbox{.}(2022)]%
        {Raffel2022}
\bibfield{author}{\bibinfo{person}{Colin Raffel}, \bibinfo{person}{Noam
  Shazeer}, \bibinfo{person}{Adam Roberts}, \bibinfo{person}{Katherine Lee},
  \bibinfo{person}{Sharan Narang}, \bibinfo{person}{Michael Matena},
  \bibinfo{person}{Yanqi Zhou}, \bibinfo{person}{Wei Li}, {and}
  \bibinfo{person}{Peter~J. Liu}.} \bibinfo{year}{2022}\natexlab{}.
\newblock \showarticletitle{Exploring the Limits of Transfer Learning with a
  Unified Text-to-Text Transformer}.
\newblock \bibinfo{journal}{\emph{The Journal of Machine Learning Research}}
  \bibinfo{volume}{21}, \bibinfo{number}{1} (\bibinfo{year}{2022}),
  \bibinfo{pages}{5485--5551}.
\newblock


\bibitem[Rajpurkar et~al\mbox{.}(2016)]%
        {rajpurkar2016squad}
\bibfield{author}{\bibinfo{person}{Pranav Rajpurkar}, \bibinfo{person}{Jian
  Zhang}, \bibinfo{person}{Konstantin Lopyrev}, {and} \bibinfo{person}{Percy
  Liang}.} \bibinfo{year}{2016}\natexlab{}.
\newblock \bibinfo{title}{SQuAD: 100,000+ Questions for Machine Comprehension
  of Text}.
\newblock
\newblock
\showeprint[arxiv]{1606.05250}~[cs.CL]
\urldef\tempurl%
\url{https://arxiv.org/abs/1606.05250}
\showURL{%
\tempurl}


\bibitem[Soldaini and Lo(2023)]%
        {peS2o}
\bibfield{author}{\bibinfo{person}{Luca Soldaini} {and} \bibinfo{person}{Kyle
  Lo}.} \bibinfo{year}{2023}\natexlab{}.
\newblock \bibinfo{booktitle}{\emph{{peS2o (Pretraining Efficiently on S2ORC)
  Dataset}}}.
\newblock \bibinfo{type}{{T}echnical {R}eport}. \bibinfo{institution}{{Allen
  Institute for AI}}.
\newblock
\newblock
\shownote{ODC-By, \url{https://github.com/allenai/pes2o}}.


\bibitem[{UNESCO} and {International Task Force on Teachers for Education
  2030}(2024)]%
        {unesco2024global}
\bibfield{author}{\bibinfo{person}{{UNESCO}} {and}
  \bibinfo{person}{{International Task Force on Teachers for Education 2030}}.}
  \bibinfo{year}{2024}\natexlab{}.
\newblock \bibinfo{booktitle}{\emph{Global Report on Teachers: Addressing
  Teacher Shortages and Transforming the Profession}}.
\newblock \bibinfo{publisher}{{UNESCO}}, \bibinfo{address}{Paris}. 187 pages.
\newblock
\showISBNx{978-92-3-100655-5}
\urldef\tempurl%
\url{https://doi.org/10.54675/FIGU8035}
\showDOI{\tempurl}
\newblock
\shownote{CC BY-SA 3.0 IGO}.


\bibitem[Vachev et~al\mbox{.}(2022)]%
        {vachev2022leaf}
\bibfield{author}{\bibinfo{person}{Kristiyan Vachev}, \bibinfo{person}{Momchil
  Hardalov}, \bibinfo{person}{Georgi Karadzhov}, \bibinfo{person}{Georgi
  Georgiev}, \bibinfo{person}{Ivan Koychev}, {and} \bibinfo{person}{Preslav
  Nakov}.} \bibinfo{year}{2022}\natexlab{}.
\newblock \showarticletitle{Leaf: Multiple-choice question generation}. In
  \bibinfo{booktitle}{\emph{Proc.~of the European Conf. on Information
  Retrieval}}.
\newblock


\bibitem[Wang et~al\mbox{.}(2024)]%
        {wang2024llmseducation}
\bibfield{author}{\bibinfo{person}{Shen Wang}, \bibinfo{person}{Tianlong Xu},
  \bibinfo{person}{Hang Li}, \bibinfo{person}{Chaoli Zhang},
  \bibinfo{person}{Joleen Liang}, \bibinfo{person}{Jiliang Tang},
  \bibinfo{person}{Philip~S. Yu}, {and} \bibinfo{person}{Qingsong Wen}.}
  \bibinfo{year}{2024}\natexlab{}.
\newblock \bibinfo{title}{Large Language Models for Education: A Survey and
  Outlook}.
\newblock
\newblock
\showeprint[arxiv]{2403.18105}~[cs.CL]
\urldef\tempurl%
\url{https://arxiv.org/abs/2403.18105}
\showURL{%
\tempurl}


\bibitem[Wang et~al\mbox{.}(2018)]%
        {wang2018qgnet}
\bibfield{author}{\bibinfo{person}{Z. Wang}, \bibinfo{person}{A.~S. Lan},
  \bibinfo{person}{W. Nie}, \bibinfo{person}{A.~E. Waters},
  \bibinfo{person}{P.~J. Grimaldi}, {and} \bibinfo{person}{R.~G. Baraniuk}.}
  \bibinfo{year}{2018}\natexlab{}.
\newblock \showarticletitle{QG-Net: A Data-Driven Question Generation Model for
  Educational Content}. In \bibinfo{booktitle}{\emph{Proceedings of the Fifth
  Annual ACM Conference on Learning at Scale}} (26--28 June 2018).
  \bibinfo{address}{London, UK}.
\newblock


\bibitem[Yadav et~al\mbox{.}(2023)]%
        {yadav2023cp}
\bibfield{author}{\bibinfo{person}{Gautam Yadav}, \bibinfo{person}{Ying-Jui
  Tseng}, {and} \bibinfo{person}{Xiaolin Ni}.} \bibinfo{year}{2023}\natexlab{}.
\newblock \showarticletitle{Contextualizing problems to student interests at
  scale in intelligent tutoring system using large language models}.
\newblock \bibinfo{journal}{\emph{arXiv preprint arXiv:2306.00190}}
  (\bibinfo{year}{2023}).
\newblock


\bibitem[Yudelson et~al\mbox{.}(2013)]%
        {Yudelson13}
\bibfield{author}{\bibinfo{person}{Michael~V. Yudelson},
  \bibinfo{person}{Kenneth~R. Koedinger}, {and} \bibinfo{person}{Geoffrey~J.
  Gordon}.} \bibinfo{year}{2013}\natexlab{}.
\newblock \showarticletitle{Individualized Bayesian Knowledge Tracing Models}.
  In \bibinfo{booktitle}{\emph{Proc.~of Artificial Intelligence in Education}},
  \bibfield{editor}{\bibinfo{person}{H.~Chad Lane}, \bibinfo{person}{Kalina
  Yacef}, \bibinfo{person}{Jack Mostow}, {and} \bibinfo{person}{Philip Pavlik}}
  (Eds.).
\newblock


\bibitem[Zhang and Rettinger(2014)]%
        {ZhangRettinger2014}
\bibfield{author}{\bibinfo{person}{L. Zhang} {and} \bibinfo{person}{A.
  Rettinger}.} \bibinfo{year}{2014}\natexlab{}.
\newblock \bibinfo{booktitle}{\emph{Final Ontological Word-Sense Disambiguation
  Prototype}}.
\newblock \bibinfo{type}{Deliverable} D3.2.3. \bibinfo{institution}{xLike
  Project}.
\newblock


\bibitem[Zhang et~al\mbox{.}(2021)]%
        {zhang2021review}
\bibfield{author}{\bibinfo{person}{Ruqing Zhang}, \bibinfo{person}{Jiafeng
  Guo}, \bibinfo{person}{Lu Chen}, \bibinfo{person}{Yixing Fan}, {and}
  \bibinfo{person}{Xueqi Cheng}.} \bibinfo{year}{2021}\natexlab{}.
\newblock \showarticletitle{A Review on Question Generation from Natural
  Language Text}.
\newblock \bibinfo{journal}{\emph{ACM Trans. Inf. Syst.}} \bibinfo{volume}{40},
  \bibinfo{number}{1}, Article \bibinfo{articleno}{14} (\bibinfo{date}{sep}
  \bibinfo{year}{2021}), \bibinfo{numpages}{43}~pages.
\newblock
\showISSN{1046-8188}
\urldef\tempurl%
\url{https://doi.org/10.1145/3468889}
\showDOI{\tempurl}


\bibitem[Zhang et~al\mbox{.}(2020)]%
        {zhang2020bertscore}
\bibfield{author}{\bibinfo{person}{Tianyi Zhang}, \bibinfo{person}{Varsha
  Kishore}, \bibinfo{person}{Felix Wu}, \bibinfo{person}{Kilian~Q. Weinberger},
  {and} \bibinfo{person}{Yoav Artzi}.} \bibinfo{year}{2020}\natexlab{}.
\newblock \bibinfo{title}{BERTScore: Evaluating Text Generation with BERT}.
\newblock
\newblock
\showeprint[arxiv]{1904.09675}~[cs.CL]
\urldef\tempurl%
\url{https://arxiv.org/abs/1904.09675}
\showURL{%
\tempurl}


\end{thebibliography}

\appendix
%
%
\end{document}